\documentclass[aps,prl,epsf,twocolumn,showpacs,floatfix,preprintnumbers,amsmath,amssymb,superscriptaddress]{revtex4}
\usepackage{graphicx}

\input{epsf}
\epsfclipon

\newcommand{\beq}{\begin{equation}}
\newcommand{\eeq}{\end{equation}}
\newcommand{\beqn}{\begin{eqnarray}}
\newcommand{\eeqn}{\end{eqnarray}}
\newcommand{\bearr}{\begin{array}}
\newcommand{\enarr}{\end{array}}

\newcommand{\toref}[1]{\mbox{(\ref{#1})}}

\newcommand{\eps}{\varepsilon}

\begin{document} 

\title{Characterizing dynamics with covariant Lyapunov vectors}
      
\author{F. Ginelli}
\affiliation{Service de Physique de l'Etat Condens\'e,
CEA-Saclay, 91191 Gif-sur-Yvette, France}

\author{P. Poggi}
\affiliation{Dipartimento di Fisica, INFN and CSDC, 
Universit\'a di Firenze, via G. Sansone 1, 50019 Sesto Fiorentino, Italy}

\author{A. Turchi}
\affiliation{Dipartimento di Fisica, INFN and CSDC, 
Universit\'a di Firenze, via G. Sansone 1, 50019 Sesto Fiorentino, Italy}

\author{H. Chat\'e}
\affiliation{Service de Physique de l'Etat Condens\'e,
CEA-Saclay, 91191 Gif-sur-Yvette, France}

\author{R. Livi}
\affiliation{Dipartimento di Fisica, INFN and CSDC, 
Universit\'a di Firenze, via G. Sansone 1, 50019 Sesto Fiorentino, Italy}

\author{A. Politi}
\affiliation{ISC-CNR via Madonna del Piano 10, 50019 Sesto Fiorentino, Italy}

\begin{abstract}
A general method to determine covariant Lyapunov vectors in both discrete-
and continuous-time dynamical systems is introduced. This allows to address 
fundamental questions such as the degree of hyperbolicity, 
which can be quantified in terms of the transversality of these intrinsic vectors.
For spatially extended systems, 
the covariant Lyapunov vectors have localization properties 
and spatial Fourier spectra qualitatively different
from those composing the orthonormalized basis obtained in the standard procedure
used to calculate the Lyapunov exponents.
\end{abstract}

\maketitle

Measuring Lyapunov exponents (LEs) is a central issue in the investigation
of chaotic dynamical systems because they are intrinsic observables that
allow to quantify a number of different physical properties such as sensitivity
to initial conditions, local entropy production and attractor
dimension \cite{review}. Moreover, in the context of spatiotemporal chaos, the
very existence of a well-defined Lyapunov spectrum in the thermodynamic limit
is a proof of the extensivity of chaos \cite{ext_chaos}, 
and it has been speculated that the small exponents 
contain information on the ``hydrodynamic'' modes of the dynamics 
(e.g., see \cite{ecpo} and references therein).

In this latter perspective, a growing interest has been devoted not only to the
LEs but also to some corresponding vectors, with the motivation
that they could contribute to identifying both the real-space structure of 
collective modes \cite{Demonte} and the regions characterized by stronger/weaker
instabilities \cite{egolf}.
However, 
the only available approach so far is
based on the vectors yielded by the 
standard procedure used to calculate the LEs \cite{benettin}.
This allows to identify the most expanding subspaces, but has the drawback
that these vectors ---that we shall call Gram-Schmidt vectors (GSV) after the 
procedure used---
are, by construction, orthogonal, even where stable and unstable manifolds are
nearly tangent. Moreover, GSV are not invariant under time reversal, and they are not
covariant, i.e. the GSV at a given
phase-space point are not mapped by the linearized dynamics into the GSV
of the forward images of this point.

While the existence, for invertible dynamics, 
of a coordinate-independent, local decomposition of phase
space into covariant Lyapunov directions ---the so-called Oseledec splitting 
\cite{review}---
has been discussed by Ruelle long ago \cite{Ruelle}, it received almost no
attention in the literature, because of the absence of algorithms to
practically determine it.
In this Letter, we propose an innovative approach based on both forward and backward
iterations of the tangent dynamics, which allows determining a set of
directions at each point of phase space that are invariant under time
reversal and covariant with the dynamics. We argue that, for any invertible
dynamical system, the intrinsic tangent space decomposition introduced by these 
covariant Lyapunov vectors (CLV) coincides with the Oseledec
splitting.


As a first important and general application of the CLV, we show that they
allow to quantify the degree of hyperbolicity of the dynamics.
Considering that all physically relevant dynamical systems are not
hyperbolic (i.e. stable and unstable manifolds are not everywhere transversal),
and that many of the available theoretical results have been derived 
under the assumption of strict hyperbolicity
(a prominent example being the Gallavotti-Cohen
fluctuation theorem~\cite{gallavotticohen}),
it is indeed highly desirable to
develop a tool to quantify deviations from hyperbolicity.
At the moment, this is doable only in very simple systems such as the H\'enon map
or the Duffing oscillator, where homoclinic tangencies can be detected by
iterating separately the tangent dynamics forward and backward in time. 
Since CLV correspond to the local
expanding/contracting directions, we can straightforwardly evaluate their
relative transversality and, accordingly, quantify the degree of hyperbolicity.
Note that GSV, being  mutually orthogonal, are useless in this context.
In a second important application of CLV
we show that, contrary to the weak localization of GSV,
they are generically localized in physical space,
providing an intrinsic, hierarchical decomposition of spatiotemporal chaos. 
Furthermore, the
knowledge of CLV paves the way to analytical methods for determining the
LEs as ensemble- rather than time-averages.

{\it Description of the algorithm.} We first summarize 
the standard method for computing the LEs 
(we consider, for simplicity, a $N$-dimensional discrete-time dynamical system).
Let ${\bf x}_{n-1}\in \mathcal R^N$ denote the phase-space point at time $t_{n-1}$
and let $\{{\bf g}_{n-1}^j\}$, $j=1, \ldots N$,   
be the $N$ orthogonal vectors obtained by applying the Gram-Schmidt
orthogonalization procedure to $N$ tangent-space vectors (we shall call this the
$(n-1)$th GS basis). Iterating the evolution equations once,
${\bf g}_{n-1}^j$ is transformed into
$\overline {\bf g}_n^j ={\bf J}_{n-1} {\bf g}_{n-1}^j$, where ${\bf J}_n$ is the
Jacobian of the transformation evaluated at time $t_n$. 
The $n$th GS basis is thereby obtained by applying the Gram-Schmidt transformation to the
vectors $\overline {\bf g}_n^j$. This amounts to computing the so-called QR
decomposition of the matrix 
$\overline{\bf G}_n = (\overline {\bf g}_n^1 | \ldots | \overline {\bf g}_n^N)$
whose columns are the Jacobian-iterated vectors of the $(n-1)$th GS basis:
$\overline{\bf G}_n = {\bf Q}_n {\bf R}_n$.
The $n$th GS basis is given by the columns of the orthogonal matrix 
${\bf Q}_n$, while ${\bf R}_n$ is an upper-triangular matrix whose
off-diagonal nonzero elements are obtained by projecting each vector
$\overline {\bf g}_n^j$ onto the subspace spanned by $\{\overline {\bf g}_n^k\}$
with $k<j$. It has been shown \cite{ershovpotapov} that, by repeating the above
procedure up to a time $t_m$ for $m$ much larger than $n$, the GS basis
converges to an orthogonal set of vectors $\{{\bf e}_m^k\}$, $k=1,\ldots, N$ -
the $m$th Gram-Schmidt vectors - which solely depend on the phase space
point ${\bf x}_{m}$.

The LEs $\lambda_1 \geq \lambda_2 \geq \ldots \geq \lambda_N$
are then nothing but the time-averaged values of the logarithms of the diagonal
elements of ${\bf R}_n$. The method we propose also exploits the usually disregarded 
information contained in the off-diagonal elements. 
Let us now assume that a set of GSV has been generated by iterating the generic
initial condition ${\bf x}_0$. Let ${\bf u}_m^j$ be a generic
vector inside the subspace $S_m^j$ spanned by $\{{\bf e}_m^k\}$, $k=1,\ldots, j$,
i.e. the first $j$ GSV at time $t_m$. 
We now iterate this vector backward in time by inverting the
upper-triangular matrix ${\bf R}_m$: 
if the $c_m^{ij}=({\bf e}_m^i \cdot {\bf u}_m^j )$ 
are the coefficients expressing it in terms of the GSV in ${\bf x}_m$, one has
$c_{m-1}^{ij}= \sum_k [{\bf R}_m]^{-1}_{ik} c_m^{kj}$, where $[{\bf R}]_{ij}$ 
is a matrix element of ${\bf R}$.  
Since ${\bf R}_m$ is upper-triangular, it is easy to verify that
${\bf u}_n^j \in S_n^j$ at all times $t_n$. This is due to the fact
that $S_n^j$ is a covariant subspace. 
Iterating ${\bf u}_m^j$
backward for a sufficiently large number $(m-n)$ of times, it eventually
aligns with the (backward) most expanding direction within ${\bf S}_n^j$. 
This defines  ${\bf v}_n^j$, our intrinsic 
$j$-th (forward) expanding direction at the phase-space point ${\bf x}_n$. 
It is straightforward to verify that ${\bf v}_n^j$ is covariant.
Define the matrix $[{\bf C}_m]_{ij} = c_m^{ij}$; then
one has ${\bf C}_m = {\bf R}_m {\bf C}_{m-1}$. By multiplying both sides
by ${\bf Q}_m$ and substituting $\overline {\bf G}_m$ for its
QR decomposition on the resulting right hand side,
one is simply left with ${\bf v}_{m}^j={\bf J}_{m-1} {\bf v}_{m-1}^j$
for $j=1,\ldots, N$. 
The CLV are independent of
where the backward evolution is started along a given trajectory, provided that
it is sufficiently far in the future. Moreover, we have verified that they are 
invariant under time reversal, i.e. that the
direction of ${\bf v}_n^j$ is the same whether we first move backward along a
given trajectory (applying the standard orthonormalization procedure) and then
forward (according to the above outlined methodology).

Our CLV $\{{\bf v}_{m}^k\}$ thus constitute an intrinsic, covariant basis defining 
expanding/contracting directions in phase space \cite{NOTE}.
The LEs are simply obtained from the CLV: the $i$th exponent is the
average of the growth rate of the $i$th vector \cite{yanchuk}.
We have checked on simple invertible maps that
they coincide with the Oseledec splitting in ${\bf x}_m$.
We conjecture that this is the case for any  invertible system. 
Note that our CLV are also well defined
for non-invertible dynamics, since it is necessary and sufficient
to follow backward a trajectory previously generated forward in time.
In this respect they provide an extension of the Oseledec splitting.
Finally, and retrospectively, a
preliminary evidence of the validity of our approach was given in
\cite{PTL}, where CLV were introduced to characterize time periodic
orbits in a 1D lattice of coupled maps. There, it was 
found that the
number of nodes (changes of sign) in a CLV is directly connected to the
position of the corresponding LE within the Lyapunov
spectrum.


We stress that the determination of the CLV can be very efficient,
making them a truly 
practical tool (as opposed, say, to calculating directly
the Oseledec splitting in the case of invertible dynamics). Indeed,
the major computational bottleneck is the memory required to
store the matrices ${\bf R_n}$ and the $n$-time GSV during the forward 
integration. This difficulty
can be substantially reduced by occasionally storing the instantaneous
configuration in real and tangent space and re-generating the rest when needed.

{\it Numerical analysis.} 
We measured the CLV in four one-dimensional systems made of $L$
nonlinear units coupled to their nearest neighbors. 
Periodic boundary conditions are used. 
The first is a chain of chaotic tent maps (TM) on the unit interval,
\beq
\begin{array}{ll}
x_{n+1}^i& = (1-2\eps) f(x_n^i) + \eps \left[f(x_n^{i+1})+f(x_n^{i-1})\right]\\ 
&\\
&{\rm with}\; f(x) = ax \;\;\;\;\;\;\;{\rm if}\; x\leq1/a \\
&{\rm and} \;\;  f(x)=\frac{a (x-1)}{1-a} \;\;{\rm otherwise.}
\end{array}
\label{TM}
\eeq
In the following we fix $\eps=0.2$ and $a=2.3$.

The second system is a chain of symplectic maps (SM),
\beq
\begin{array}{l}
p_{n+1}^i = p_n^i + \mu \left[g(q_n^{i+1}-q_n^{i})-g(q_n^{i}-q_n^{i-1})\right] \\
q_{n+1}^i = q_{n}^i + p_{n+1}^i 
\end{array}
\label{SM}
\eeq
where $g(z) = \sin(2 \pi z)/(2 \pi)$. This model was studied in \cite{RadonsMap}
to analyse the so-called ``hydrodynamic Lyapunov modes''. 
Eq.~(\ref{SM}) conserves total momentum $P=\sum_i p^i$, and is invariant under a
translation of the $q$ coordinates. Therefore, the Lyapunov spectrum 
possesses two null exponents. In the following we fix $\mu=0.6$. 

The last two models are second-order continuous-time systems governed by
\beq
\ddot q_i = F(q_{i+1}-q_{i}) - F(q_{i}-q_{i-1})\;.
\eeq
For $F(x) = \sin(x)$, we
have the rotator model (RM), while for $F(x) = x + x^3$, the system reduces to
a Fermi Pasta Ulam chain (FPU). These two widely studied Hamiltonian systems
provide a good testing ground to investigate the connection between microscopic
dynamics and statistical mechanics. 
Besides the zero LE associated with a shift along the trajectory, 
both models have three other
null LEs arising from energy and momentum conservation plus translational
invariance. 
Numerical simulations have been performed at energy
density $E/L=1$ (for the RM) and $E/L=10$ (for FPU).

\begin{figure}
\includegraphics[draft=false,clip=true, width=8.6cm]{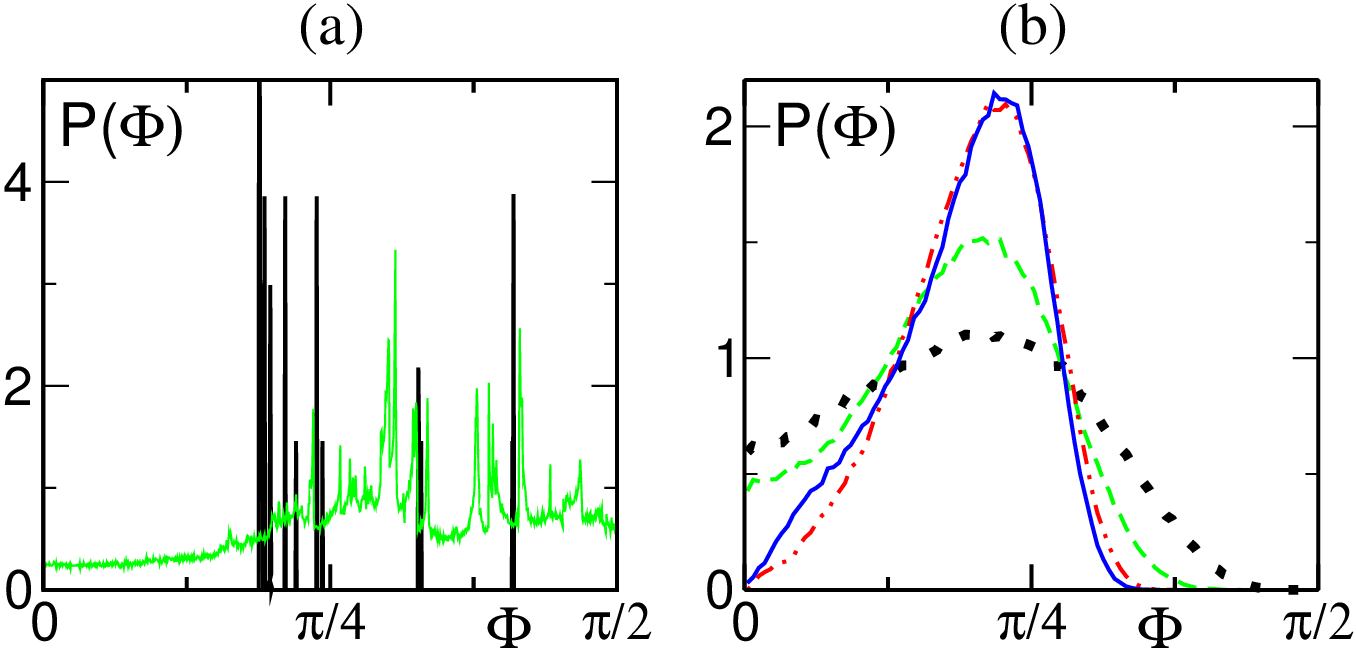}
\caption{
(Color online).
Probability distribution of the angle between stable and unstable manifold. 
$(a)$ H\'enon map $x_{n+1} = 1 -1.4\, x_n^2 + 0.3 x_{n-1}$ (green light line),
and Lozi map $x_{n+1} = 1 -1.4\,|x_n| + 0.3 x_{n-1}$ (black line, rescaled by a 
factor 10). $(b)$  
TM ($L=12$, black dotted line), SM ($L=10$, green dashed line), RM ($L=32$, red dot-dashed line),
and FPU ($L=32$, blue full line).}
\label{Hyp}
\end{figure}

{\it Hyperbolicity.} A dynamical system is said to be hyperbolic if its phase
space has no homoclinic tangencies, i.e. the stable and unstable manifolds are
everywhere transversal to each other. In the mathematical literature, it is
known that the Oseledec splitting is connected to hyperbolicity~\cite{Bochi}, but 
the lack of practical algorithms to determine the splitting makes such results
of little use in physically relevant contexts. Here, the knowledge of the
CLV allows testing hyperbolicity by determining the angle 
between each pair $(j,k)$ of expanding ($j$) and contracting ($k$) directions
\begin{equation}
\phi_n^{j,k} = \cos^{-1}(|{\bf v}_n^j \cdot{\bf v}_n^k|) \in [0,\pi/2]
\end{equation}
where the absolute value is taken because signs are irrelevant. As a first test,
we have computed the probability distribution $P(\phi)$
of $\phi_n^{1,2}$ for two classic two-dimensional maps. 
Arbitrarily small angles are found for the H\'enon map, 
while the distribution is bounded away from zero in the Lozi map
(Fig.~\ref{Hyp}a). This is perfectly consistent with the 
well-known fact
that only the latter model is hyperbolic \cite{Collet}.

In spatially extended systems, given the multi-dimensional character of the
invariant manifolds, it is appropriate to determine the minimum angle,
$\Phi_n = \min \{\phi_n^{j,k} |( {\bf v}_n^j\in E_n^{+}, {\bf v}_n^k\in E_n^{-})\}$
where $E_n^{\pm}$ are the expanding and contracting invariant subbundles at time
$t_n$ along the trajectory. The histograms in Fig.~\ref{Hyp}b 
show that models \toref{TM} and \toref{SM}  
are characterized by stronger hyperbolicity violations than the Hamiltonian systems.
Altogether, recalling that $\Phi$ refers to the least transversal pair
of directions, we are led to conclude that the dynamics of high-dimensional systems 
should be closer to hyperbolic than that of low-dimensional ones. 
This justifies the often-made assumption that spatially-extended systems 
are practically hyperbolic.

\begin{figure}
\includegraphics[clip, width=8.6cm]{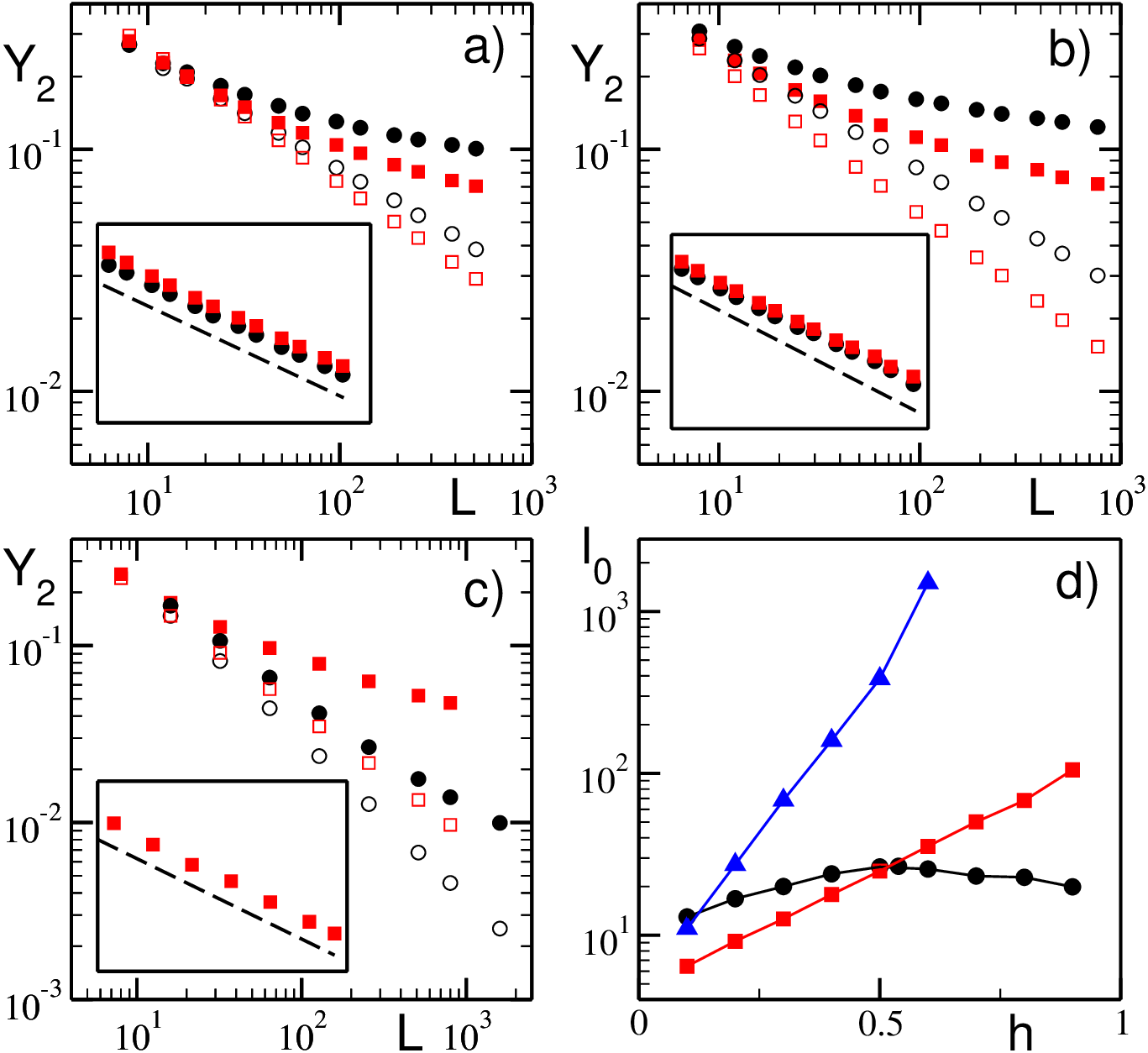}
\caption{(Color online).
Inverse participation ratio $Y_2$ (see text) 
of both CLV and GSV for different dynamics. 
Time averages were performed over typically $10^5 \sim 10^6$ timesteps
and cubic splines have been employed to interpolate $Y_2(h, N)$ between the discrete set of values
$h$, $j=1,\ldots,N$.   
$(a-c)$: Log-log plot of $Y_2$ as a function of chain length $L$ 
at fixed spectrum position $h$.
CLV results are shown in full symbols, while
GSV by empty symbols. In the log-log scale insets: inverse of the localization length 
$\ell$ has been subtracted from $Y_2$ to better show
the CVL behavior $Y_2(L) \sim 1/\ell + L^{-\gamma}$ (see text). The dashed black lines mark
a decay as $\gamma = \frac{1}{2}$.  
$(a)$: TM for  
$h=0.1$ (black circles) and $h=0.4$ (red squares).
$(b)$: SM 
for $h=0.2$ (black circles) and $h=0.4$ (red squares).
$(c)$: FPU ($h=0.2$, black circles) and RM ($h=0.2$, red squares).
$(d)$: Lin-log plot of the asymptotic 
localization length $\ell$ of CLV as a function of $h$ for 
TM (black circles) SM (red squares) and RM (blue triangles).}
\label{Localization}
\end{figure}

{\it Localization properties in extended systems}. The spatial structure of 
the vectors associated to the LEs is of interest in many contexts.
We now show that the GSV ---which have been used so far--- and the CLV
have {\it qualitatively} different localization properties. One usually considers
the inverse participation ratio \cite{part} $Y_2 = \langle \sum_i (\alpha^j_i)^4 \rangle$
where $\langle \cdot \rangle$ indicates an average over the trajectory
and $\alpha^j_i$ is a measure of the component of the $j$th vector at site
$i$ (with the normalization $\sum_i |\alpha^j_i|^2=1$). In systems characterized by a single local 
real variable (such as our TM), $\alpha^j_i$ 
is taken to be the $i$-th component of the $j$-th
CLV or GSV, while in the case of symplectic systems, where two components are
present (${\bf v}^j =(\delta {\bf q}^j,\delta{\bf p}^j)$), it is natural 
to choose $(\alpha^j_i)^2 = (\delta q^j_i)^2 + (\delta p^j_i)^2$. In
order to investigate the thermodynamic limit, it is necessary to determine
$Y_2(h,L)$ for fixed $h=(j-\frac{1}{2})/L$ and increasing $L$. On the one hand,
localized vectors are characterized by a finite inverse participation ratio,
$Y_2(h, L)\to 1/\ell$, for $L\to \infty$, where $\ell$ is a localization
``length''. 
On the other hand, in completely delocalized structures,
$Y_2(h, L) \sim 1/L$.

In Fig.~\ref{Localization} we show how $Y_2$ typically scales with the chain
length $L$. The GSV show weak (de)localization: 
their participation ratio exibits an $h$-dependent ``dimension'' $\eta(h)$:
$Y_2 \sim L^{-\eta(h)}$. One can show that this anomalous behavior is entirely due
to the Gram-Schmidt procedure, and has nothing to do with the dynamics \cite{unpublished}.
On the other hand, CLV are localized objects.
For TM, SM and RM dynamics we find good evidence of the
scaling law $Y_2(h,L) \sim  1/\ell(h) + L^{-\gamma}$ 
with $\gamma \approx \frac{1}{2}$. This allows for a reliable determination of $\ell$.
For the FPU dynamics, we find only slight curvature in the log-log plot of
Fig.~\ref{Localization}c, signalling
that larger system sizes are probably needed 
to definitely enter the scaling regime.
Moreover, for symplectic dynamics the localization length $\ell(h)$ diverges
as $h \to 1$ (Fig.~\ref{Localization}d).  
Assuming the continuity of the LE spectrum, the divergence of $\ell$ is  
not surprising, since the conservation laws imply that the Lyapunov vectors
(both GSV and CLV) corresponding to $h=1$ (i.e. to null LEs) are
completely delocalized.

\begin{figure}
\includegraphics[draft=false,clip=true, width=8.6cm]{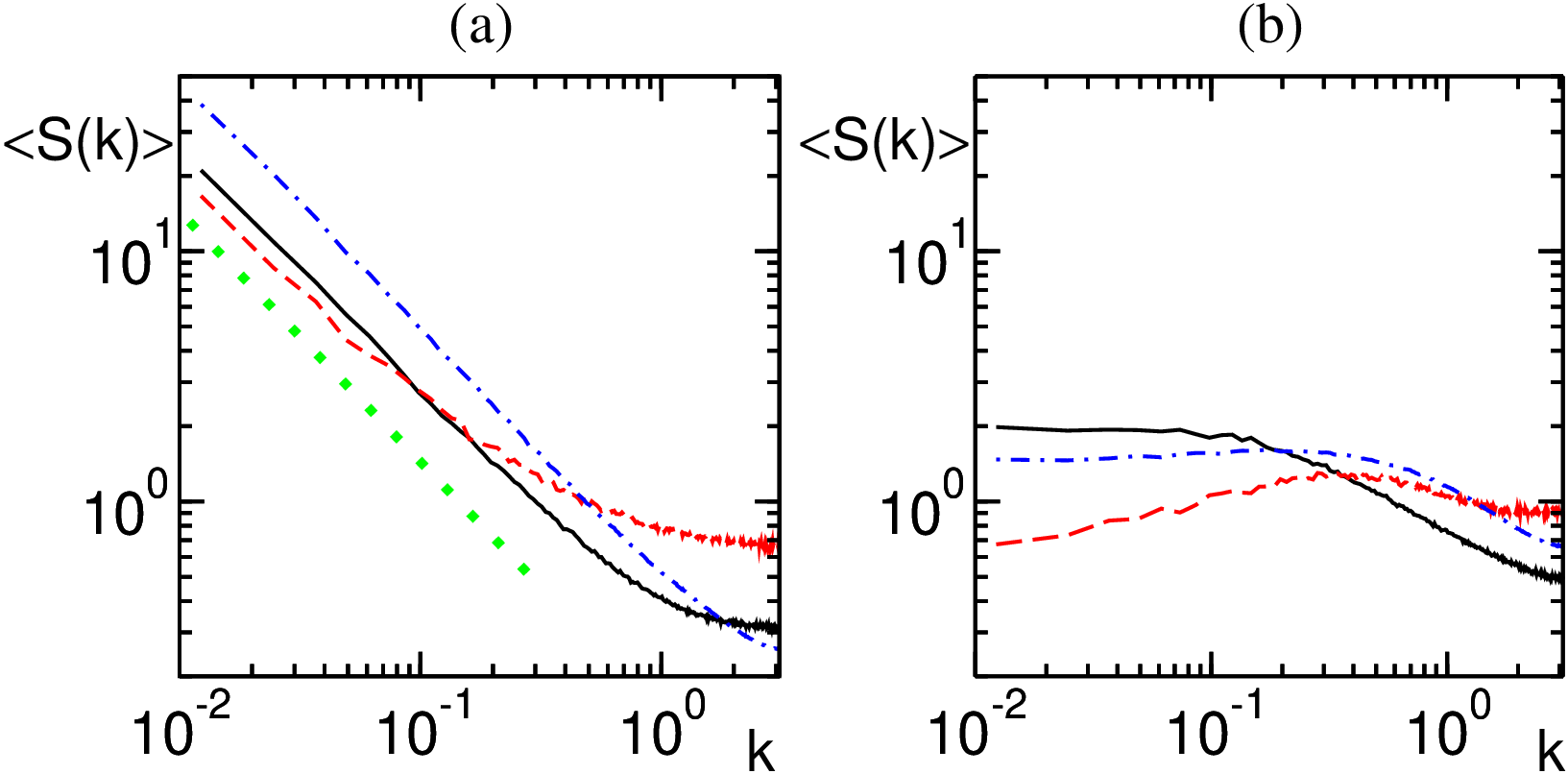}
\caption{(Color online). 
Trajectory averaged power spectrum (as a function of the wavenumber $k=j\,2\pi / L$, $j=1,\ldots, L/2$) 
of the space components of CLV $(a)$ and GSV $(b)$ corresponding to the smallest
positive LE. Solid (black), dashed (red) and dot-dashed lines (blue) refer to FPU,
RM and SM respectively ($ L = 512$).
The dotted green line, corresponding to a $1/k$ behavior is plotted for comparison in panel (a).}
\label{fouri}
\end{figure}

{\it Fourier analysis}.
Another way proposed to characterize the spatial structure of a Lyapunov vector is to
look at its power spectrum $S(k) = \left|\sum_m \beta_m {\rm
e}^{imk}\right|^2$, where $\beta_m$ denotes the vector component 
associated with the space coordinate $q_m$ at site $m$.
For instance, this was used in \cite{RadonsMap} in the context of the investigation of 
so-called ``hydrodynamic'' modes (only GSV were considered there).
Here, we have focused on the vector corresponding to the smallest positive LE
in our symplectic models, for which this LE goes continuously to zero
as the system size increases (note that GSV and CLV coincide 
for the null exponents linked to symmetries and conservation laws).
We observe again a clear qualitative difference between the spectra of GSV and CLV
(Fig.~\ref{fouri}).
In particular, the near-zero CLV exhibit an intriguing low-frequency divergence 
of the $1/k$ type in all three symplectic models we have analysed.
Thus, the qualitative difference between GSV and CLV extends to the $h\to 1$
case.

{\it Perspectives}.
Now that the local directions of stable and unstable manifolds are made
available in generic models, many questions can be addressed in a more
accurate way: Quantifying (non-)hyperbolicity in the context of the (numerical) attempts
to ``verify'' the fluctuation theorem is one. Another set of questions relates 
to the spatial structure of the dynamics in extended systems, such as
the quantification of local degree of chaos (amount of instability), a hierarchical
decomposition of spatiotemporal chaos, the search for true, intrinsic, collective
(``hydrodynamic'') modes, etc.
A further field where the knowledge of CLV can
help to make progress is optimal forecast in nonlinear models. Here the knowledge of
the local transversality of the invariant manifolds can indeed be combined with
the so-called bred vectors to use the information on the past evolution to
decrease the uncertainty along unstable directions \cite{bred}. 

\acknowledgments


\begin{thebibliography}{99}

\bibitem{review} J.-P. Eckmann and D. Ruelle, Rev. Mod. Phys. {\bf 57} 617 (1985). 
\bibitem{ext_chaos} 
D. Ruelle, {\sl Thermodynamic Formalism}, Reading, MA: Addison \& Wesley (1978);
R. Livi, A. Politi and S. Ruffo, J. Phys. A {\bf 19}, 2033 (1986);
P. Grassberger, Phys. Scri. {\bf 40}, 346 (1989).
\bibitem{ecpo} J.-P. Eckmann, C. Forster, H.A. Posch and E. Zabey, J. Stat. Phys {\bf 118}, 813 (2005).
\bibitem{Demonte} N. Nakagawa and Y. Kuramoto, Physica D {\bf 80}, 307 (1995);
S. De Monte, F. d'Ovidio, H. Chat\'e and E. Mosekilde, Phys. Rev. Lett. {\bf 92}, 254101 (2004).
\bibitem{egolf} K. Kaneko, Physica D, {\bf 23}, 436 (1986);
H. Chat\'e, Europhys. Lett. {\bf 21} 419, (1993);
D.A. Egolf, E.V. Melnikov, W. Pesch and R.E. Ecke, Nature,
{\bf 404}, 733 (2000).
\bibitem{benettin} I. Shimada and T. Nagashima, Prog. Theor. Phys. {\bf 61}, 1605 (1979);
G. Benettin, L. Galgani, A. Giorgili, and J.M. Strelcyn, Meccanica, {\bf 15}, 21 (1980).
\bibitem{Ruelle} D. Ruelle, Publ. Math. IHES {\bf 50}, 275 (1979). 
\bibitem{gallavotticohen} G. Gallavotti and E.G.D. Cohen, Phys. Rev. Lett. {\bf 74}, 2694 (1995).
\bibitem{ershovpotapov} S.V. Ershov, A.B. Potapov, Physica D {\bf 118}, 167 (1998).
\bibitem{NOTE} We expect that the
CLV are ill-defined in the presence of degeneracies; in such cases, they have
to be grouped according to the multiplicity of the corresponding LE.
\bibitem{yanchuk} As it has been shown for 
simple 2D maps
and 3D smooth flows:
B. Eckhardt, D. Yao, Phyisica D {\bf 65}, 100 (1993);
G. Froyland, K. Judd and A. I. Mees, Phys. Rev. E {\bf 51}, 2844 (1995);
A. Politi, F. Ginelli, S. Yanchuk, and Yu. Maistrenko, Physica
D, {\bf 224} 90 (2006).
\bibitem{PTL} A. Politi, A. Torcini, S. Lepri, J. Phys. IV
{\bf 8}, 263 (1998).
\bibitem{RadonsMap} H. Yang and G. Radons, Phys. Rev. E {\bf 73}, 016202 (2006).
\bibitem{Bochi} Ya. B. Pesin, Russian Math. Surveys {\bf 32}, 55 (1977);
J. Bochi, M. Viana, Ann. I.H. Poincar\'e {\bf 19}, 113 (2002).
\bibitem{Collet} More precisely, the Lozi map is hyperbolic with the exception of a 
zero measure set of cuspidal points where the tangent bundle is not defined:
P. Collet and Y. Levy, Commun. Math. Phys. {\bf 93}, 461 (1984).
\bibitem{part} A. D. Mirlin, Phys. Rep. {\bf 326}, 259 (2000).
\bibitem{unpublished} J. Kockelkoren and H. Chat\'e, {\sl unpublished}.
\bibitem{bred} D. Patil {\it et al.}, Phys. Rev. Lett. {\bf 86}, 5878 (2001).
\end{thebibliography}
\end{document}